\documentclass{llncs}
\usepackage{graphicx}
\newtheorem{fig}{{Figure}}
\newtheorem{thm}{{Theorem}}
\newtheorem{tab}{{Table}}
\newtheorem{algo}{{Algorithm}}
\newtheorem{hypo}{{Assumption}}

\def\ligne#1{\hbox to \hsize{#1}}

\def\grostrait{\hbox to\hsize{\vrule height 1pt depth 1pt width \hsize}}
\def\demitrait{\hbox to\hsize{\vrule height 0.5pt depth 0.5pt width \hsize}}
\def\ttV{\vrule height 10pt depth 4pt width 0.6pt}
\def\ttH{\hrule height 0.3pt depth 0.3pt width \hsize}

\begin{document}

\begin{center}
{\bf \Large
A protocol for a message system for the tiles of the heptagrid, in the hyperbolic
plane}
\vskip 20pt
Maurice {\sc Margenstern}
\vskip 20pt

professor emeritus of Universit\'e Paul Verlaine $-$ Metz\\
LITA, EA 3097, UFR-MIM, and CNRS, LORIA\\
Campus du Saulcy, 57045 Metz, C\'edex, France\\
{\em e-mail}: {\tt margens@univ-metz.fr} 
\end{center}

\begin{abstract}
This paper introduces a communication system for the tiles of the
heptagrid, a tiling of the hyperbolic plane. The method can be extended to
other tilings of this plane. The paper focuses on an actual implementation
at the programming stage with a short account of two experiments.
\end{abstract}

{\em Keywords: hyperbolic tilings, cellular automata,
applications.
}

\normalsize \vfill

\section{\Large Introduction}
\label{introduction}

In this paper, we present a protocol to manage communications between tiles
of a tiling of the hyperbolic plane.
Why the hyperbolic plane? Because the geometry
of this space allows to implement there any tree structure. The reason for that
is that the simplest tilings defined in this plane are spanned by a tree. 
We refer to the author's papers and books on this topic, see~\cite{mmbook1,mmbook2}. 
Tree structures are already intensively used in computer science, especially by
operating systems. 
But the fact that trees are naturally embedded in this geometrical space, 
especially on tilings living there, was never used. 
This paper proposes to take advantage of this property. 

   In Section~\ref{geom}, we remind what is needed of hyperbolic geometry in
order the reader could understand the content of the paper. 

In Section~\ref{geom}, we remind what is needed of hyperbolic geometry in
order the reader could understand the content of the paper. In
Section~\ref{navigation}, we sketchily describe the navigation technique
with a new aspect which was not used in~\cite{mmLNEEhongkong}. In
Section~\ref{scenario}, we present the protocol which allows us to improve
the system briefly mentioned in~\cite{mmLNEEhongkong}. 
In Section~\ref{program}, we give an account of the simulation program and
in Section~\ref{experiment} we present the experiment which was performed by
running the simulation program. In Section~\ref{conclusion}, we
conclude with further possible development of the scenario implemented by the 
protocol.

\section{\Large Hyperbolic geometry and its tilings}
\label{geom}
\noindent

Our first sub-section reminds the Poincar\'e's disc model which allows us to
have a partial visualization of the hyperbolic plane. Our second sub-section defines
the simplest tilings which can there be defined, allowing us to construct
{\bf grids}.
In our third sub-section, we focus on the tiling on which our proposal is based,
the tiling $\{7,3\}$ of the hyperbolic plane which we call the {\bf heptagrid}.

\subsection{Hyperbolic geometry}
\label{hypgeom}
\noindent
Hyperbolic geometry appeared in the first half of the
19$^{\rm th}$ century, proving the independence of the parallel
axiom of Euclidean geometry. This was the end of a search during two thousand
years in order to prove that the well known axiom about parallels in Euclid's
treatise is a consequent of his other axioms. The search is by itself a very
interesting story, full of deep teachings of high price for the philosophy of
sciences, we recommend the interested reader to have a look at~\cite{bonola}.
Hyperbolic geometry was the first issue of the notion of axiomatic independence.
But also, it raised the first doubts on the absolute power of our abstract mind.
It opened the way to the foundational works of mathematical logics, so that
computer science, the daughter of logics, appears to be a grand-child of
hyperbolic geometry. The paper hopes to show that this kind of relations
holds not only on a philosophical ground.

   The search failed. This was proved in the second third of the 19$^{\rm th}$
century by the discovery of hyperbolic geometry. In this new geometry, all the
non-parallel axioms of Euclidean old and, from a point~$A$ out of a line~$\ell$ 
and in the plane defined by~$\ell$ and~$A$ there are two parallels to~$\ell$
passing through~$A$. Around forty years after the discovery, models of the
new geometry in the Euclidean one were found. Presently, we turn to the most popular
model of the hyperbolic plane nowadays, Poincar\'e's model.

\subsection{The Poincar\'e's disc model}
\label{poincare}
\noindent

This model is represented by Figure~\ref{poincare_disc}.
   In the model, the points of the hyperbolic plane are represented by the points
of an open disc fixed in advance called the {\bf unit disc}. The circle which
is the boundary of the disc is called the set of {\bf points at infinity} and we
denote it by~$\partial U$. These
points do not belong to the hyperbolic plane, but they play an important role
in this geometry. The lines of the hyperbolic plane are represented by the trace
in the disc of its diameters or by the trace in the disc of circles which are orthogonal
to~$\partial U$. It is not difficult to see that the lines are also characterized in the
models by their points at infinity as diameters of the disc and circles orthogonal
to~$\partial U$ meet $\partial U$ twice exactly.
\vskip 7pt
\vtop{
\centerline{
\mbox{\includegraphics[width=150pt]{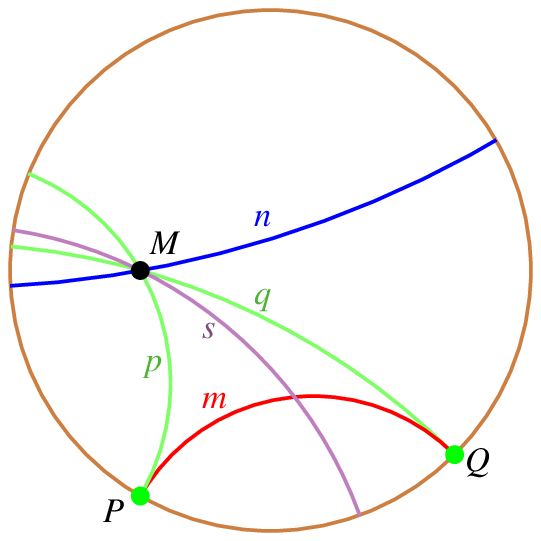}}
}
\begin{fig}\label{poincare_disc}
\small
\hskip -5pt :
Poincar\'e's disc model. We can see that in the model, both lines~$p$ and~$q$ 
pass through~$A$ and that they are parallel to~$\ell$: $p$ and~$q$ touch~$m$ in 
the model at~$P$ and~$Q$, points at infinity.
\end{fig}
}
\vskip 7pt
   The figure also shows us an important feature of the new plane: outside parallel
and secant lines to a given line~$\ell$, passing through a point~$A$ not on~$\ell$,
there are also lines passing through~$A$ which do not cut~$\ell$, neither inside the disc,
nor on~$\partial U$ and nor outside the disc: they are called {\bf non-secant} lines.
A last but not least property of the new plane is that there are no similarity: a figure
of this plane cannot be re sized at another scale. Resizing necessarily changes the
shape of the figure.

\subsection{The tilings $\{p,q\}$ of the hyperbolic plane}
\label{pq}
\noindent
   Consider the following process. We start from a convex regular polygon~$P$. We 
replicate $P$~in its sides and, recursively, the images in their sides. If we cover 
the plane without overlapping, then we say that $P$ {\bf tiles the plane by 
tessellation}. We shall often say for short that  $P$ {\bf tiles the plane}, here, 
the hyperbolic
plane. A theorem proved by Poincar\'e in 1882 tells us that if~$P$ has $p$~sides and 
if its interior angle is $\displaystyle{{2\pi}\over q}$, then $P$ tiles the hyperbolic
plane, provided that:
\vskip 7pt
\hbox to \hsize{\hfill$\displaystyle{1\over p}+\displaystyle{1\over q} < 
\displaystyle{1\over 2}$.
\hfill(1)} 
\vskip 7pt
\noindent
The numbers $p$ and~$q$ characterize the tiling which is denoted $\{p,q\}$ and
the condition says that the considered polygons live in the hyperbolic plane.
This inequality entails that
there are infinitely many tessellations of the hyperbolic plane. The tessellation
attached to $p$ and~$q$ satisfying the inequality is denoted by $\{p,q\}$.

Note that we find the well known Euclidean tessellations
if we replace $<$ by~$=$ in the above expression.
We get, in this way, $\{4,4\}$ for the square, $\{3,6\}$ for the
equilateral
triangle and $\{6,3\}$ for the regular hexagon.

   In~\cite{mmbook1,mmbook2}, the author provides the reader with a uniform treatment
of these tessellations. The basic feature is that each tessellations is spanned by
a tree whose structure obeys well defined properties. We shall illustrate these
points in our next sub-section where we focus on a particular tessellation: the
tiling $\{7,3\}$ of the hyperbolic plane which we call the {\bf heptagrid}. 

\subsection{The heptagrid}
\label{hepta}

The heptagrid is illustrated by Figure~\ref{til73}.
The figure is very symmetric, but at this stage, it is difficult to
identify each tile of the figure, especially for the tiles which are outside the
first two rings around the central cell.

The tiles look very much the hexagonal tiles of the corresponding tessellation of the
Euclidean plane,but the global organization is very different. It seems to us that
this figure indicates that we need something to locate the tiles. We turn to this point
in Section~\ref{navigation}

\vskip 7pt
\vtop{
\centerline{
\mbox{\includegraphics[width=160pt]{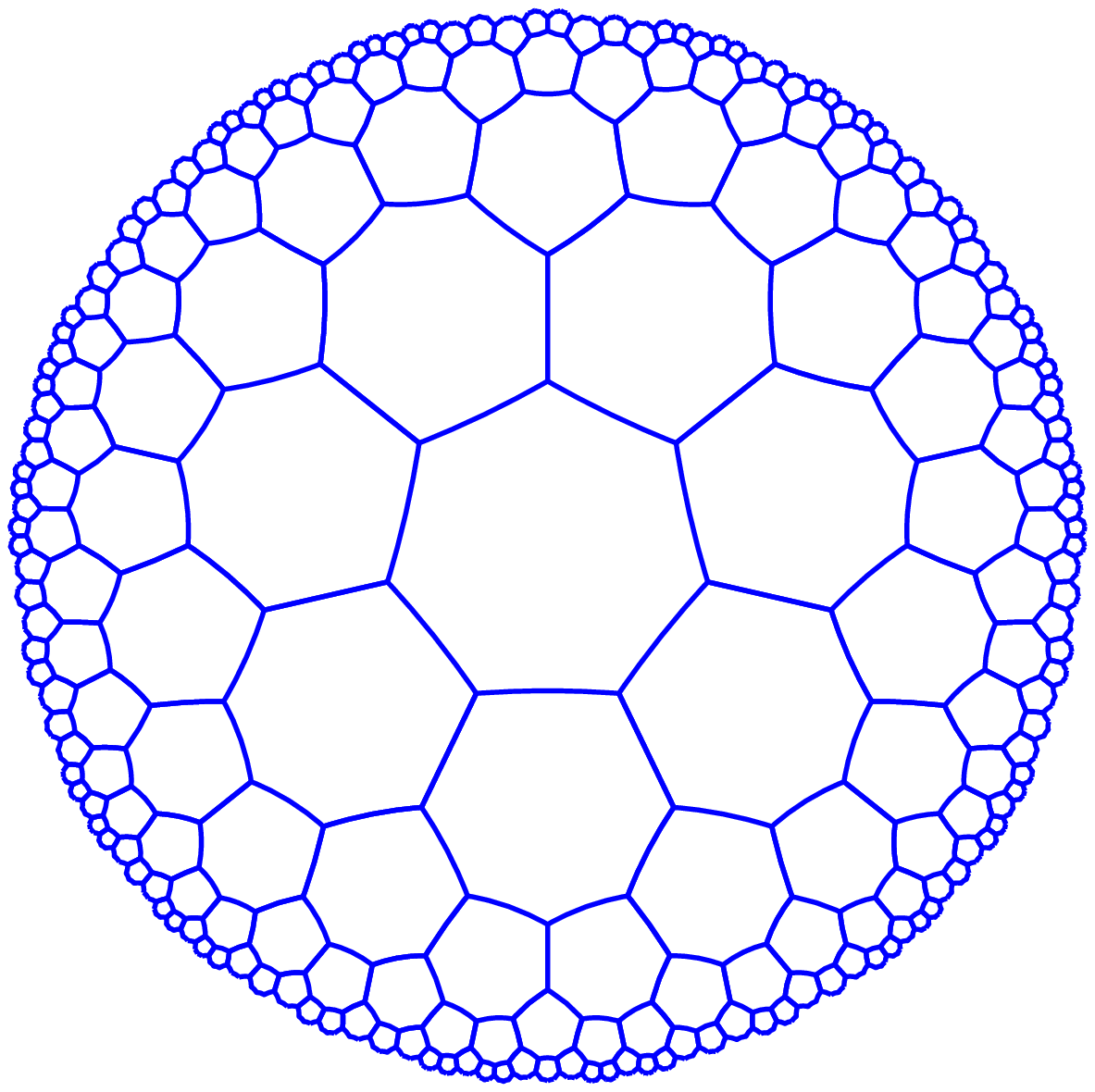}}
}
\begin{fig}\label{til73}
\small
The heptagrid: an illustrative representation.
\end{fig}
}
\vskip 7pt
   
\section{\Large Navigation in the heptagrid}
\label{navigation}

   We have seen on the heptagrid that it is not easy to locate the tiles of this tiling,
especially as we go further and further from the centre of the disc.

   This is the point to draw the attention of the reader on two points. The disc model
may be misleading if we forget that we are in the Euclidean plane and that some symmetries
of the figure have no counter part in hyperbolic geometry. As an example, the central tile
seems to play an important role, in particular its centre. However, the hyperbolic plane
has no central point as well as the Euclidean plane has no such point. We decide to
fix an origin when we define coordinates in the Euclidean plane. This is the same here.
The central tile is simply a tile which we decide to be the origin of our coordinate system.
The second point is that Poincar\'e's disc model gives us a local picture only. We see
the immediate neighbourhood of a point which we decided to place at the centre of the disc.
The models looks like a small window on the hyperbolic plane whose central part only
is well observable. From this lens effect, we conclude that walking on the hyperbolic
plane, we are in the situation of the pilot of a plane flying with instruments only.

    And so, what are our instruments?

The first two pictures of Figure~\ref{tree_73} represent them. On the left-hand
side of the figure, we can seen the {\bf mid-point lines} defined by the
following property: the line which joins the mid-points of two consecutive sides
of a heptagon cuts a heptagon of the tiling at the mid-point of a side only.
If we consider two rays issued from the meeting point of two secant mid-point
lines and defined by the smallest angle, we can define a set of tiles
for which all vertices but possibly one or two lie inside this angle. We call this
the restriction of the tiling to a sector. The remarkable property is that the
restriction of the tiling to a sector is spanned by a tree, the tree which is
illustrated by the right-hand side of the figure. In the middle of the figure,
we can see that the whole tiling can exactly be split into a central cell and
seven sectors dispatched around the central tile.

\vskip 7pt
\vtop{
\centerline{\hskip 10pt
\mbox{\includegraphics[width=100pt]{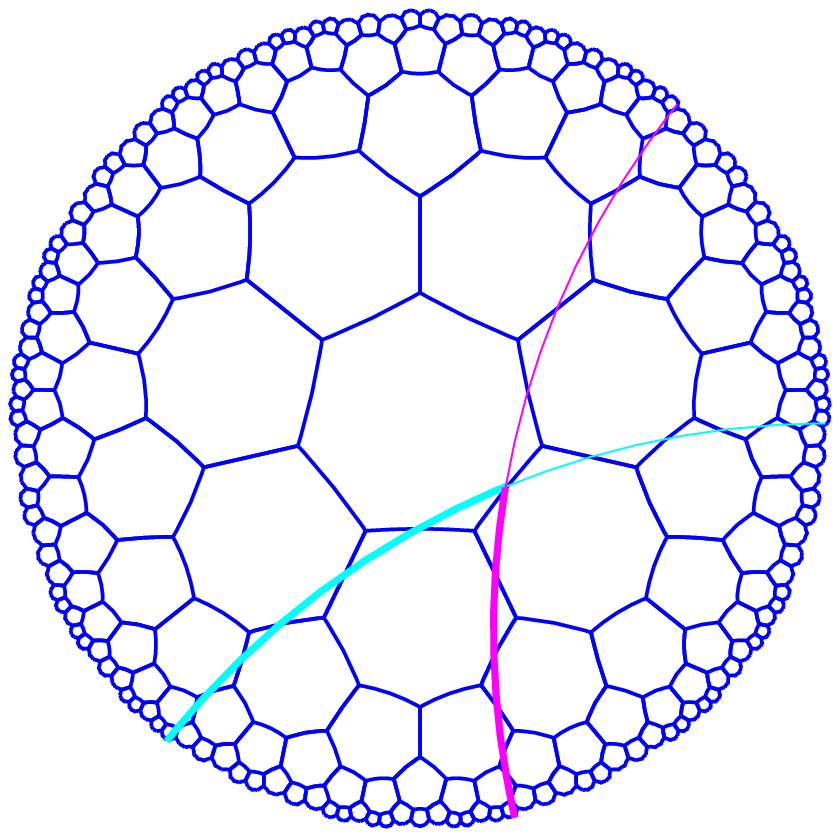}}
\hskip 10pt
\raise2.5pt\hbox{\mbox{\includegraphics[width=97.5pt]{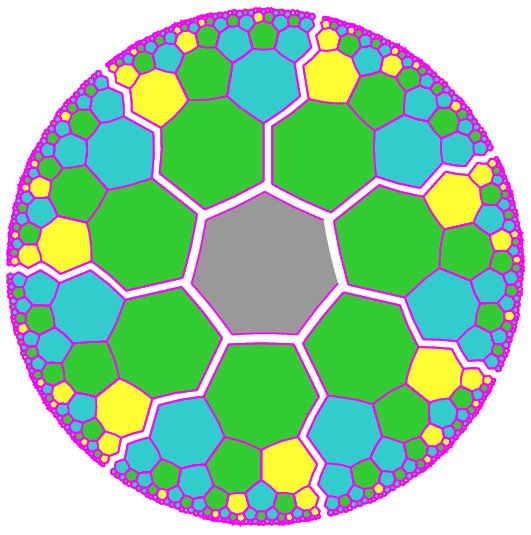}}}
\raise-2.5pt\hbox{\mbox{\includegraphics[width=130pt]{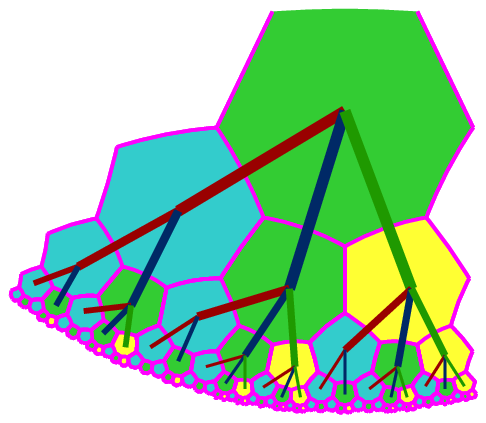}}}
}
\begin{fig}\label{tree_73}
\small
Left-hand side: the mid-point lines. This tool which shows how a sector spanned
by the tree is defined in the heptagrid.
\vskip 0pt
Middle: first part of the splitting, around a central tile, fixed in advance,
seven sectors. Each of them is spanned by the tree represented in the right-hand
side.
\vskip 0pt
Right-hand side: the tree which spans the tiling.
\end{fig}
}
\vskip 7pt

The tree which spans the tiling can be generated by very simple rules
indicating the exact connection between a node and its sons. There are two
kind of nodes, the {\bf black} and the {\bf white} ones. The rules are,
in self-explaining notations:
\vskip 7pt
\hbox to\hsize{\hfill
$B\rightarrow BW$,\hskip 20pt $W\rightarrow BWW$
\hfill(2)}
\vskip 7pt
\noindent
From these rules, it can be proved that the
number of nodes which are on the level~$n$ of the tree is $f_{2n+1}$ where
$\{f_n\}_{n\in I\!\!N}$ is the Fibonacci sequence defined by $f_0=f_1=1$ and
the induction equation \hbox{$f_{n+2}=f_{n+1}+f_n$}. For this reason, the
tree is called the {\bf Fibonacci tree}.

It is known that natural numbers
can be represented as sums of distinct terms of the Fibonacci sequence:
$n=\displaystyle{\sum\limits_{i=0}^ka_if_i}$, with $a_i\in\{0,1\}$, $a_k\not=0$ if
$n\not=0$. This representation is not unique, but it can be made unique by requiring
$k$~to be maximal in the above representation. This is the {\bf greedy Fibonacci}
representation of~$n$. It is characterized by the fact that considering the $a_i$'s
in their order as a word on $\{0,1\}^*$, there are no contiguous~1's in the word.

Presently, number the nodes of the tree level by level and, on each level, from the
left to the right, starting from the root which receives~1 as its number. Next,
call {\bf coordinate} of a node~$\nu$ the greedy Fibonacci representation of
its number. Then the coordinates of the nodes of the tree have this striking
property. If $[\nu]$ is the coordinate of~$\nu$, among the sons of~$\nu$ there
is a single node whose coordinate is~$[\nu]00$, which is called the
{\bf preferred son}. Also, we can rewrite~(2) as~(3)
\vskip 7pt
\hbox to\hsize{\hfill
$B\rightarrow \overline{B}W$,\hskip 20pt $W\rightarrow B\overline{W}W$
\hfill(3)}
\vskip 7pt
\noindent
where the bar indicates the position of the preferred son.

A {\bf path} between a tile~$A$ and a tile~$B$ is a sequence $\{T_i\}_{i\in[0..n]}$
with $T_0=A$, $T_n=B$ and, for each $i\in[1..n]$, $T_{i-1}$ and~$T_i$ have a
common side. We say that $n$~is the length of the path and we note that from
this definition, the path is oriented: its {\bf source} is the tile $T_0$ and its
{\bf target} is the tile~$T_n$. We also say that the path goes from~$A$ to~$B$.
Clearly, a path from~$B$ to~$A$ is obtained by reversing the numbering of the $T_i$'s
defining the path from~$A$ to~$B$.

For complexity reasons, it is convenient to take a {\bf shortest} path between~$A$
and~$B$: it is a path between~$A$ and~$B$ whose length is minimal. Note that in general
there is no unique shortest path but, by the very definition, such a path exists.
An important particular case is when both tiles are on the same branch of a tree: 
this part of the branch is a shortest path between the tiles.

From the properties of the preferred son, we obtain:

\begin{thm}\label{pathlin} {\rm(see\cite{mmASTC,mmbook1})}
There is an algorithm which computes the path from a node of the Fibonacci tree 
to the root from the coordinate of the node which is linear in the size of the 
coordinate.
\end{thm}

From this algorithm, it is easy to compute a path between two nodes which is most
often almost a shortest path between the nodes: first, go from the nodes to the
root and connect the two paths at the root. Moreover, the computation is linear
in the size of the two coordinates. However, in certain situations, this is not
the shortest way. Now, in~\cite{mmbook2}, we proved a refinement of
Theorem~\ref{pathlin}. In order to
state it properly, we have to define coordinates for the tiles of the heptagrid.

To do this, we look at the middle picture of Figure~\ref{tree_73}. We
increasingly number the sectors from~1 up to~7 by counter-clockwise turning around
the central cell, fixing the sector which receives number~1 once and for all. Now,
the coordinate of a tile is defined by~0 for the central tile and, for any other
tile~$T$, by the couple~$(\sigma,\nu)$ where $\sigma\in\{1..7\}$ defines the sector
which contains the tile and where~$\nu$ is the coordinate of~$T$ in the tree which
spans the sector.

Now, we can state the result:

\begin{thm}\label{strongpathlin} {\rm(see\cite{mmbook2})}
There is an algorithm which computes a shortest path between two tiles of the
heptagrid which is linear in the size of the coordinates of the tiles.
\end{thm}

   Another consequence of Theorem~\ref{pathlin} is the computation of the 
coordinates of the neighbours of a tile, where a neighbour of the tile~$T$ is a
tile~$N$ which shares a side with~$T$. Clearly, $T$ is also a neighbour of~$T$
and we shall often say that~$N$ and~$T$ are neighbours.

   The computation of the coordinates of the neighbours relies on two functions of~$n$,
the number of a tile: $f(n)$ which is the number of the father of the tile and 
$\sigma(n)$ which is the number of the preferred son of the tile. According to the
previous notations, it is also interesting to consider the function~$[f]$ such that
\hbox{$[f]([n]) = [f(n)]$} and the function $[\sigma]$ such that
\hbox{$[\sigma]([n]) = [\sigma(n)]$}. From the proof of Theorem~\ref{pathlin},
there is an algorithm which allows to compute $[f]$ and~$[\sigma]$ which is linear in
the size of~$[n]$. However, note that in the computation of the path the application
of the algorithm at one step is in fact a constant. With the help of this function,
the coordinates of the neighbours of a tile are given by Tables~\ref{neigha}
and~\ref{neighb}.
In the table, the neighbours of a tile~$T$ are increasingly numbered from~1 to~7
while counter-clockwise turning around~$T$. Neighbour~1 is the father of~$T$. The 
father of a root is the central tile. For the central tile, the neighbour~$i$ is 
the tile associated to the root of the tree in the sector~$i$. If~$i$ is the number 
of a neighbour of~$T$, we say that the side shared by~$T$ and this neighbour is 
also numbered by~$i$.

\def\lignetaba #1 #2 #3 {%
\hbox to \hsize{%
\ttV\hbox to 25pt{\hfill#1\hfill}
\ttV\hbox to 50pt{\hfill#2\hfill}
\ttV\hbox to 50pt{\hfill#3\hfill}\ttV
}
}
\vtop{
\begin{tab}\label{neigha}
\rm\small
\hskip -5pt :
The numbers of the neighbours for a tile~$\nu$ which is inside the tree. The tile
may be black or white. 
\end{tab}
\hbox to\hsize{\hfill
\vtop{\offinterlineskip\hsize=133pt
\ttH
\lignetaba {$\tau$} {black} {white}
\ttH
\ttH
\lignetaba 1 {$f(\nu)$} {$f(\nu)$} 
\ttH
\lignetaba 2 {$f(\nu)$$-$1} {$\nu$$-$1} 
\ttH
\lignetaba 3 {$\nu$$-$1} {$\sigma(\nu)$$-$$1$}
\ttH
\lignetaba 4 {$\sigma(\nu)$} {$\sigma(\nu)$} 
\ttH
\lignetaba 5 {$\sigma(\nu)$+1} {$\sigma(\nu)$+1}
\ttH
\lignetaba 6 {$\sigma(\nu)$+2} {$\sigma(\nu)$+2}
\ttH
\lignetaba 7 {$\nu$+1} {$\nu$+1} 
\ttH
}
\hfill}
\vskip 12pt
}

Table~\ref{neigha} considers the general case, when the node associated to the
tile is inside the tree. This means that the corresponding node always has its 
father in the tree and that all its neighbours are also in the tree. In 
Table~\ref{neighb}, we have the exceptional cases. The nodes on the 
leftmost branch, the root excepted, and those which are on the rightmost branch, 
the root excepted. The nodes of the rightmost branch are white and the root is 
also white. The nodes on the leftmost branch, the root excepted, are black.
 
 It remains to indicate that in the case of a heptagon~$H$ which is on the
left- or the rightmost branch, it is easy to define the number of the sector
to which belongs the neighbours which do not belong to the tree of~$H$.
Indeed, let $\sigma$~be the number of the sector in which $H$~lies. If $H$~is
a black node, its neighbours~2 and~3 are in the sector
$\sigma\ominus1$, where $\sigma\ominus1=\sigma$$-$$1$ when $i>1$ and
$1\ominus1=7$. If $H$~is a white node, its neighbours~6 and~7
are in the sector $\sigma\oplus1$ with $\sigma\oplus1=\sigma$+$1$ when $i<7$
and $7\oplus1=1$. Note that for the root of the
sector~$\sigma$, its neighbour~2 is in the sector $\sigma\ominus1$ and its
neighbour~1 is the central
cell which is outside all the sectors.

\def\lignetabb #1 #2 #3 #4 {%
\hbox to \hsize{%
\ttV\hbox to 25pt{\hfill#1\hfill}
\ttV\hbox to 50pt{\hfill#2\hfill}
\ttV\hbox to 50pt{\hfill#3\hfill}
\ttV\hbox to 50pt{\hfill#4\hfill}\ttV
}
}
\vtop{
\begin{tab}\label{neighb}
\rm\small
\hskip -5pt :
The numbers of the neighbours for a tile~$\nu$ which is either the root of the tree,
or which belongs to an extremal branch the leftmost or the rightmost ones. The numbers
are given in the columns {\tt root}, {\tt left} and {\tt right} respectively. 
\end{tab}
\hbox to \hsize{\hfill
\vtop{\leftskip 0pt\offinterlineskip\hsize=185.6pt
\ttH
\lignetabb {$\tau$} {left} {right} {root}
\ttH
\ttH
\lignetabb 1 {$f(\nu)$} {$f(\nu)$} 0
\ttH
\lignetabb 2 {$\nu$$-$1} {$\nu$$-$1} 1
\ttH
\lignetabb 3 {$\sigma(\nu)$$-$$1$} {$\sigma(\nu)$$-$$1$} {$\sigma(\nu)$$-$$1$}
\ttH
\lignetabb 4 {$\sigma(\nu)$} {$\sigma(\nu)$} {$\sigma(\nu)$}
\ttH
\lignetabb 5 {$\sigma(\nu)$+1} {$\sigma(\nu)$+1} {$\sigma(\nu)$+1}
\ttH
\lignetabb 6 {$\sigma(\nu)$+2} {$\nu$+1} {$\nu$+1}
\ttH
\lignetabb 7 {$\nu$+1} {$f(\nu)$+$1$} 1
\ttH
}
\hfill}
\vskip 7pt
}

Both Theorems~\ref{pathlin} and~\ref{strongpathlin} give an algorithm to
compute new coordinates if we change the place of the central tile to another tile.
In Section~\ref{program}, we briefly give an explicit pseudo code implementing
an algorithm satisfying Theorem~\ref{strongpathlin}.

The computation of the functions $f$ and~$\sigma$ used in the computation of the
coordinates of the neighbours of a tile are also used for this purpose and based
on these considerations and on the previous results we have that:

\begin{thm}\label{chgcoord} {\rm(see~\cite{mmbook2})}
Consider a system of coordinate and a tile~$T$. Assume that we take~$T$ as the central
tile and that we fix its new side~$1$. There is an algorithm which, for each tile~$T'$
computes the coordinates of~$T'$ in the new system in linear time in the size of the
coordinate in the initially given system. 
\end{thm}

   It is important to remark that in order to obtain the linearity of the algorithm,
we do not compute the functions~$f$ and~$\sigma$ but $[f]$ and~$[\sigma]$ applied
to~$[n]$. This means that in order to compute $\sigma(n)$$-$$1$ for instance,
we need an algorithm for computing $[m$$-$$1]$ from~$[m]$. A similar remark holds for
$[m$+$1]$. The needed algorithms can be found in~\cite{mmbook1}.

\section{\Large The communication protocol}
\label{scenario}

   In Section~\ref{introduction}, we mentioned that in~\cite{mmLNEEhongkong}, 
we already proposed a communication protocol for the tiles of the heptagrid.
This protocol was based on a specific system of coordinates, inherited 
from~\cite{mmJCAcomm,mmbook2}. For the convenience of the reader, we briefly
describe this system in Sub-section~\ref{absolrel}. Then, in
Sub-section~\ref{protocol} we define the new protocol.

\subsection{Absolute and relative systems}
\label{absolrel}

The {\bf absolute} system is based on a numbering of the sides tiles of the
heptagrid. For each tile, we number the sides from~1 to~7{} in this order while
counter-clockwise turning around the tile. Now, how to fix side~1? We again take
the situation of the left-hand side of Figure~\ref{tree_73}: a central cell surrounded
by seven sectors, each one spanned by a copy of the Fibonacci tree. Now, side~1 is
fixed once and for all for the central tile. For the other tiles, side~1 is the
side shared by the tile and its father, considering that the central cell is the
father of the root for each copy of the Fibonacci tree spanning the sectors.

Now, a side always belongs to two tiles and so it receives two numbers. 
This is why this numbering is called {\bf local}. However, the association
between both numbers is not arbitrary. There is a correspondence between them 
although it is not one to one. When one number is known, the status of the tile 
and the fact that the corresponding node whether lies or not on an extremal 
branch of the tree are also needed to determine the other number.
This correspondence is given by Table~\ref{sidenumbers} and Table~\ref{thepairs} 
lists all the couples used by the sides: note that we are far from using all 
possible couples.

   The absolute system consists in first fixing the local numbering once and for all.
Table~\ref{sidenumbers} will still be used to determine the two numbers of a side
of a heptagon.  

\def\lignetabii #1 #2 #3 #4 {%
\hbox to \hsize{%
\ttV\hbox to 40pt{\hfill#1\hfill}
\ttV\hbox to 40pt{\hfill#2\hfill}
\ttV\hbox to 40pt{\hfill#3\hfill}
\ttV\hbox to 40pt{\hfill#4\hfill}\ttV
}
}
\vtop{
\begin{tab}\label{sidenumbers}
\rm\small
\hskip -5pt :
Correspondence between the numbers of a side shared by two heptagons,
$H$ and~$K$. Note that if $H$~is white, the other number of side~$1$
may be~$4$ or~$5$ when $K$~is white and that it is always~$5$ when
$K$~is black.
\end{tab}
\hbox to \hsize{\hfill
\vtop{\offinterlineskip\hsize=170.6pt
\ttH
\hbox to\hsize{\ttV\hfill\hbox to 80pt{\hfill black $H$\hfill}
\hfill\ttV\hfill
\hbox to 80pt{\hfill white $H$\hfill}\hfill\ttV}
\ttH\ttH
\lignetabii {in $H$} {in $K$} {in $H$} {in $K$}
\ttH\ttH
\lignetabii 1 {3$^{wK}, $4$^{bK}$} 1 {4$^{wK}$, 5}
\ttH
\lignetabii 2      6     2     7
\ttH
\lignetabii 3      7     3     1
\ttH
\lignetabii 4      1     4     1
\ttH
\lignetabii 5      1     5     1
\ttH
\lignetabii 6      2     6     2
\ttH
\lignetabii 7      2     7 {2$^{wK}$, 3$^{bK}$}
\ttH
\vskip 7pt
}
\hfill}
}

Next, we remark that the local numbering gives a way to encode a path between
two tiles~$A$ and~$B$. Let $\{T_i\}_{i\in[0..n]}$ be a shortest path from~$A$
to~$B$ and denote by $s_i$ the side shared by $T_i$ and~$T_{i+1}$ for 
$i\in[0..n$$-$$1]$. Let $a_i$ be the number of $s_i$ in~$T_i$ and $b_i$ be its
number in $T_{i+1}$. Then we say that the sequence $\{(a_i,b_i)\}_{i\in[0..n-1]}$ 
is an {\bf address} of~$B$ {\bf from}~$A$. The reverse sequence gives an
address of~$A$ from~$B$. However, on the just above sequence,
we do not know from the sequence itself that $(a_n,b_n)$ is the last side. 
In order to do this, we change a bit the association of the numbers: for $T_i$
belonging to the path, we denote by~$en_i$ the side shared with $T_{i-1}$ and
by~$ex_i$ the side shared with~$T_{i+1}$. For $T_0$, as $en_0$ cannot be defined
as the number of a neighbour of~$T_0$, we put $en_0=0$. This time we say that
the sequence $\{(en_i,ex_i)\}_{i\in[O..n]}$ is the {\bf coordinate} of $B$ from~$A$.
Similarly, the sequence $\{(ex_{n-i},en_{n-i})\}_{i\in[0..n]}$ is the coordinate
of~$A$ from~$B$. And so, the correspondence between the address and the coordinate
is easy: $a_i = ex_i$ and $b_i = en_{i+1}$ for $i\in[0..n$$-$$1]$ which contains
the definitions of~$ex_0$ and~$en_n$.

   Now, how to define a shortest path between $A$ and~$B$?

   There are two ways: the first way is given by Theorem~\ref{strongpathlin}. We
apply the algorithm defined in Section~\ref{program} in order to find the
coordinate of $A$ from~$B$.

   The second way consists in the following. If $A$~sends messages to every tile,
it considers itself as the central tile, taking its own number~1 as the number~1
of the central cell. Remember that all tiles have the same size, the same shape and
the same area, this is why each tile may feels 'equal' to the others. When it sends 
the message to its neighbours, it also send them
the information that it is the central cell and it sends $(0,i)$ to its 
neighbour~$i$. And so, the neighbour receives its coordinate from~$A$.
By induction, we assume that each tile~$T$ which receives the message from~$A$
also receives its coordinate from~$A$ and its status in the relative tree to~$A$
in which $T$~is. From this information, and as~$T$ knows from which
neighbour it receives the message, $T$ know which of its neighbours are its relative
sons and so, it can append the element $(en_T,ex_T)$ to the address it conveys to
the corresponding son together with the relative status of the son. And so,
we proved that each tile receiving a message from~$A$ also receives its address 
from~$A$ and its relative status with respect to~$A$. In fact, we have an 
implementation
of the local numbering attached to~$A$ as a central tile. This local numbering is
called the {\bf relative} system. Now, from its coordinate from~$A$, $T$~may
computes the coordinate of~$A$ from~$T$ in a linear time in the size of the 
coordinate, as follows from what we already have noticed. And so, if
$T$~wishes to reply to the message sent by~$A$ it can do it easily.
Moreover, from the properties we have seen in Section~\ref{navigation},
we can see that, proceeding in the just described way, a public message is sent
to every tile once exactly, which is an important feature.

\subsection{The protocol}
\label{protocol}
\noindent
We have now the tools to describe the protocol of communication between the tiles.

For this protocol, we distinguish two types of messages, {\bf public} ones and
{\bf private} ones. By definition, a public message is a message sent by a tile to all
the other tiles. A private message is a message sent by a tile to a single other one.
This distinction belongs to the sender of the message.
In this protocol, we assume that we have a global clock defining a discrete
time and that a message leaving a tile~$T$ at time~$t$ can reach only a neighbour
of~$T$ at time~$t$+1. We say that the maximal speed for a message is~1.
 
   The public message makes use of the relative system of the sender. 
However,
in the coordinates which are constructed by the tiles which relay the message,
the numbers $en_i$ and~$ex_i$ computed by the relaying tile are defined according
to the absolute system as the tile does not know where is the sender and
as its own local numbering is defined by the absolute system. 

   A private message is either a reply to a message, either public or private,
or a message sent to a single tile according to the following procedure. Each
tile~$T$ has a direct access to a the system. Given the coordinates of a tile~$N$
as defined in Section~\ref{navigation}, the central cell being that of the absolute
system, the managing system gives to~$T$ a shortest path from~$T$ to~$N$ 
which is a coordinate of~$N$ from~$T$. And so, a private message is defined by
the fact that it has the address of the receiver. 

\def\lignetab #1 #2 {%
\hbox to \hsize{%
\ttV\hbox to 30pt{\hskip 10pt#1\hfill}
\ttV\hbox to 40pt{\hfill #2\hfill}\ttV
}
}
\vtop{
\begin{tab}\label{thepairs}
\small
\hskip -5pt :
The pairs $(i,j)$ of numbers of a side of a heptagon. It is assumed that
the first number denotes the side 
\end{tab}
\vspace{-10pt}
\hbox to\hsize{\hfill
\vtop{\offinterlineskip\hsize=74.1pt
\ttH
\lignetab 1 {(1,3)}
\lignetab {} {(1,4)}
\lignetab {} {(1,5)}
\ttH
\lignetab 2 {(2,6)}
\lignetab {} {(2,7)}
\ttH
\lignetab 3 {(3,7)}
\lignetab {} {(3,1)}
\ttH
}
\hfill
\vtop{\offinterlineskip\hsize=74.1pt
\ttH
\lignetab 4 {(4,1)}
\ttH
\lignetab 5 {(5,1)}
\ttH
\lignetab 6 {(6,2)}
\ttH
\lignetab 7 {(7,2)}
\lignetab {} {(7,3)}
\ttH
}
\hfill}
\vskip 19pt
}
 
  In order to deliver the information efficiently, a private message from~$A$
to~$B$ stores the coordinate of~$B$ as two stacks~$a$ and~$r$. The stack~$a$ is 
for the direct run from~$A$ to~$B$, the stack~$b$ is for the way back. Each tile~$T$ 
on the path conveys the message to the next one~$N$ on the path, in the direction 
from~$A$ to~$B$. To perform this, $T$ reads the top of~$a$, say $(en,ex)$. It knows
that~$ex$ is the number of~$N$ from itself. Just before sending~$N$ the message and
the stacks, $T$~pops the top $(en,ex)$ of~$a$ and pushes $(ex,en)$ on the top of~$r$.
In this way, $T$ knows that it is the receiver if $ex=0$. When this is the case,
$T$~pops the top $(en,ex)$ of~$a$, pushes $(ex,en)$ on~$r$ but does nothing else. 
Note that at this moment $a$~is empty. When~$T$ is ready to answer, it exchange~$a$
and~$r$ and so, the same process allows the message to reach~$A$ together with a
coordinate of~$B$ from~$A$. This is illustrated by Figure~\ref{BobandAlice} on
a toy example.

   It is easy to see that this process is linear in time with respect to the
coordinate of the receiver, assuming that all messages travel at maximal speed~1.

   Now, we shall see that in the process of a public message, the computation
by the relaying tile of the new information which it has to append to the message
is easily computed. Here two, we have two stacks, but the stack~$r$ is always
empty.

   Consider a relaying tile~$T$. Let $(en_0,ex_0)$ be the top of the stack~$a$. By 
construction, $ex_0$~is the side of the neighbour~$N$ through which~$T$ received the
message. In order to facilitate the computation, $T$ also receives the number~$en_1$ 
in~$T$ of the side numbered~$ex_0$ in~$N$. Let~$s$ be the number of the relative son
of~$T$ in the relative tree. We know that $s\in[3..5]$ if $T$ is white in the relative
tree and that $s\in[4,5]$ if~$T$ is black in the relative tree. This index 
corresponds to a position of the father at the absolute index~1. Now, the absolute
index~$ex_1$ of the son defined by~$s$ is given by:
\vskip 7pt
\hbox to\hsize{\hfill
$ex_1 = 1 + ((en_1$$-$$1)+s$$-$$1)$ mod~7,
\hfill(4)}
\vskip 7pt
Once $ex_1$ is known, the other absolute number of the side defined by~$ex_1$,
say $en_2$, may be determined by Table~\ref{thepairs}. Now,
we know that when $ex_1\in\{1,2,3,7\}$, $en_2$~is not uniquely defined. The value
of~$en_2$ depends on the absolute status of~$T$. The simplest solution is to
assume that all pairs~$(i,j)$ for $i\in\{1..7\}$ are known for each tile~$T$
in a table~{\it output} which is a sub-table of Table~\ref{thepairs}, see
Section~\ref{program} for the implementation of this important point. Then
we have:
 
\vskip 7pt
\hbox to\hsize{\hfill
 $en_2=output(ex_1)$.
\hfill}
\vskip 7pt
For implementation, note that the root of the relative sector~1 for a sender is
its absolute neighbour~1. Also note that formula~(4) is different from the formula
given in~\cite{mmbook2}. In formula~(4), we do not need to know the relative status 
of the tile, contrarily to the formula used in~\cite{mmbook2}. This is automatically 
given by the relative indices used for the relative sons.

\vtop{
\vspace{-40pt}
\centerline{
\mbox{\includegraphics[width=200pt]{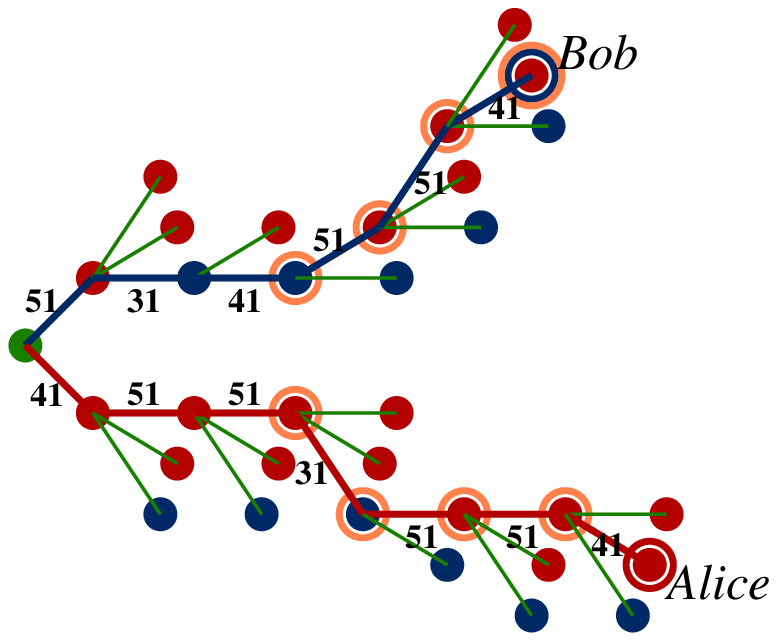}}
}
\vspace{-35pt}
\begin{fig}
\label{BobandAlice}\small
Illustration of the protocol of communication between tiles of the pentagrid or of the
heptagrid. The circled tiles indicate a shortest path from Alice to Bob. Note the
numbering of the sides shared by two tiles according to the definition of this
sub-section.
\end{fig}
}

However, we cannot assume that all tiles send messages at any time. This would not
be realistic. Also, we cannot assume that public messages are sent for ever to cover 
the whole plane which would also not be realistic. Indeed, in case of public messages
sent without stopping, the number of messages at any tile at each time would
increase to infinity at an exponential rate with time.

   In order to limit the scope of a public message, we define a {\bf radius} of
its propagation. This means that if a public message is sent from~$A$, it will
reach any tile whose distance from~$A$ is at most the radius. The distance
between two tiles~$A$ and~$B$ is the length of a shortest path between~$A$
and~$B$. Of course, the message could also bring with it the delay which could
be decremented by~1 each time it reaches a new tile, and the message would destroy
itself when the delay would be~0. The defect of this solution is that we have to 
transport the delay to all tiles within the radius and that at each time, we have
to perform this decrementing at each tile. 

   There is another solution. When $A$ sends a public message with radius~$r$,
the message is not sent at maximal speed~1 but at a speed~$\displaystyle{1\over2}$.
As we have a global clock, we shall distinguish between odd and even times. A
public message travels at odd times and remains on the tile at even times.
Consider the message~$\mu$ sent from~$A$ at time~1. Now, $A$ must remember~$\mu$
and this is implemented as another message~$\mu_e$ which is not sent immediately.
The new message has the minimal information. It has to destroy~$\mu$ and only~$\mu$.
To this purpose, all messages are identified by a unique number given by the system.
And so, $\mu_e$ contains the number of~$\mu$. Next, $A$ keeps~$\mu_e$ during
$r$~tops of the clock. In fact, $\mu_e$ also keeps a delay which is~$r$ when
$\mu_e$ is created and which is decremented at each top of the clock. When the
delay is~0, $\mu_e$ is sent to all tiles, according to the same process as a public
message, but at speed~1. Also, another important difference is that~$\mu_e$
does not need to transport a stack as it is not sent in order to get any reply.
As $\mu_e$ travels at speed~1, at time~$2r$, it reaches $\mu$ and it destroys it.
Later on we shall say that $\mu_e$ is an {\bf erasing} message.

   With this, we completed the description of the process concerning public
messages. Private messages always travel at speed~1.

   A last point for the simulation: as a tile does not send a message at any
time, we have to decide when it sends a message. For this purpose,
we use a Poisson generator, both for the decision of sending a message and, 
in the case of a public message, for defining the radius of the propagation of
the message. For each parameter, we use a different coefficient for the generator.
We shall see the values in Section~\ref{experiment}.

\section{\Large The simulation program}
\label{program}

   As usual for the implementation of a theoretical model, the simulation program
results from many choices decided by the programmer for the implementation of various
features, in particular, the structure representing the space of the simulation,
we shall see this point in Subsection~\ref{data}. Subsection~\ref{auxil}
is devoted to auxiliary computations connected with the implementation
of the basic algorithms. Subsection~\ref{scenar} describes the exact
implementation of the scenario described in Section~\ref{protocol}.
The simulation program was written in $ADA95$. When this is the case, we mention
facilities given by the programming language.

\subsection{Data structures}
\label{data}

   In a first stage of the experiments described in Section~\ref{experiment}, the
heptagrid was implemented as a table {\tt space} with two entries: one in 0..7 and the
other in 0..{\tt maxsize}, where {\tt maxsize} is the number of tiles of a sector
represented by the simulation, the sector~$i$, with $i\in[1..7]$ being represented
by {\tt space(i,*)}. From section~\ref{navigation}, it is not difficult to
compute that up to the level~$n$, the number of tiles in a sector is
$f_{2n+2}$$-$$1$.

   The elements of the table are a small table of 8 records indexed from~0 to~8.
Record~0 represents the tile {\tt space}$(i,j)$: it is the tile~$T$ of 
coordinate $(i,j)$ with $i\in[1..7]$ and
\hbox{$j\in[1..${\tt maxsize}$]$}. Record~$v$, with  $v\in[1..7]$ represents the
neighbour~$v$ of~$T$. The fields of the record contain the information needed
to minimize the computation time. Except the coordinate of the tile and of its
neighbours, the record is assumed to contain an information of constant size.

   Presently, in order to overcome a quick overflow by the manipulation of the table,
the space of the experiment is implemented by stacks. There is a basic
table {\tt space} with a single entry in 0..7 which represents the central cell
and the roots of the Fibonacci trees.
Each element of the table is a pointer, {\tt addresstile}
which points at a tile. The pointer in {\tt space(0)} points at the central cell,
{\tt space($i$)}, with \hbox{$i \in 1..7$} points at the root of the sector~$i$.
The tiles themselves are represented by a record whose fields are: 
\vskip 7pt
\vtop{\leftskip 60pt\parindent -20pt\hsize=300pt
{\tt num}: the number of the node in its tree,\vskip 0pt
{\tt sect}: the number of the sector,\vskip 0pt 
{\tt neighbour}: a table of seven pointers, {\tt neighbour($i$)} pointing
at the neighbour~$i$ of the tile,\vskip 0pt
{\tt associate}: a table of seven numbers, the absolute number of side~$i$
in the neighbour~$i$,\vskip 0pt
{\tt branch}, with the self-explaining values about the position of the tile
with respect to the borders of the sector:
\vskip 0pt 
\hskip 20pt{\tt left}, {\tt right} {\tt middle}, {\tt root}
and {\tt centre},\vskip 0pt 
{\tt status}, the status of the tile: {\tt central}, {\tt white} or {\tt black},
\vskip 0pt
{\tt outer}, boolean: says if the considered neighbour is outside the 
simulation space,\vskip 0pt
{\tt border}, boolean: says if the tile is on the border of the simulation space,
\vskip 0pt
{\tt message\_stack0}, {\tt message\_stack1}: two pointers on the stack of 
messages,\vskip 0pt
{\tt last\_message0}, {\tt last\_message1}: two pointers at the last element of 
the stack, {\it i.e.} the most recent message.
}

   The stack of messages contains what is needed for the communication system.
Each element of the stack holds a record with the following fields:

\vskip 7pt
\vtop{\leftskip 60pt\parindent -20pt\hsize=300pt
      {\tt next} : pointer for handling the stack,\vskip 0pt
      {\tt relative\_father}, an integer in 1..7, 0 if not defined,\vskip 0pt
      {\tt relative\_status}, with the values: {\tt black}, {\tt white}, {\tt centre},
\vskip 0pt
      {\tt number}, an integer: the number given to the message,\vskip 0pt
}

\vtop{\leftskip 60pt\parindent -20pt\hsize=300pt
      {\tt the\_type}, with the values: {\tt public}, {\tt private}, {\tt erasing},
\vskip 0pt
      {\tt wait}, an integer: the counter for an erasing message,
\vskip 0pt
      {\tt direct, wayback}, pointers of messages.
}
\vskip 7pt
   Now, why two pointers at the stack of messages?

   This is to perform the simulation in the following way. At each tile,
we represent the stack of messages by two disjoint stacks: {\tt stack0},
accessed through {\tt message\_stack0} and {\tt last\_message0}, and
{\tt stack1}, accessed through the pointers {\tt message\_stack1} and 
{\tt last\_message1}.
We consider that {\tt stack0} represents the stack at the tile at the time~$t$
while {\tt stack1} represents the stack at the same tile but at the
time~$t$+1. The disjunction of {\tt stack0} with respect to {\tt stack1}
avoid any confusion of pointers. And this disjunction is needed as a tile
possibly receives contributions from its neighbours and this is performed
sequentially by the simulation, whence the distinction between the two 
configurations of the same stack at the time~$t$ and at the time~$t$+1:
the same tile receives contribution from its neighbours at different steps of the
sequential process within the steps of computation which represent the turning
from the time~$t$ to the time~$t$+1.

\subsection{Auxiliary computations}
\label{auxil}
 
   The computation of the shortest path is given by a simple algorithm which
hides more involved computations, although they remain within a linear estimate
with a small coefficient, see~\ref{shortest}. Among the shortest paths from the
tile~$A$ to the tile~$B$, there are two extremal ones: if we consider the~$A$
as the central cell, $B$ lies in a sector~$i$ and the the path from $A$ to~$B$
which passes through the root of the sector and going along the branch of the
tree from the root to~$B$ is one of these extremal paths: there is no other shortest
path on the left-hand side of this path while looking at~$B$ and the root with
$B$ below the root. If we start from~$B$ as the central cell, we get the other
extremal path: no shortest to the right-hand side of this one. Now, when we have
at our disposal a shortest path~$\pi$, say from~$A$ to~$B$, it is possible to
define the leftmost shortest path from~$A$ to~$B$. This path is computed
thanks to the procedure {\tt measure} which computes the distance between
two paths. It also relies on a function {\tt pathroot} which computes the
path from a node to the root from the coordinate of the node. It uses a
function {\tt leftmost} which computes the leftmost shortest path when it
is given a shortest path.

   The function {\tt pathroot} is a significantly improved version of the
algorithm given in~\cite{mmbook2}, giving a much simpler proof of 
Theorem~\ref{pathlin}. The key point was to notice that there is a kind
of propagation of the carry when for the first time to contiguous~1's are
detected: test in the {\tt else}-branch of the main {\tt if} of the loop.

   When starting the execution of function {\tt shortest}, both paths start 
from the central cell as indicated by the instructions which initialize
{\tt Lcursor} and {\tt Rcursor}. As suggested by the identifiers, {\tt Lcursor}
points at the leftmost tiles between {\tt tile1} and {\tt tile2}. The
function {\tt theleftmost} looks whether the points belong to the same sector
of not. If not, the absolute difference between the indices of the sectors
allows us to know the leftmost tile. If they are in the same sector,
then the function follows both paths from the root to the nodes
until the paths diverge as we assume that the tiles are distinct.
At the tile which is the point where both paths diverge, it is easy to 
determine which tile is on the left-hand side of the other.

\vtop{
\begin{algo}\label{shortest}
\small
The computation of the shortest path. Given {\tt tile1} and {\tt tile2}
with the assumption that \hbox{\tt tile1 $\not =$ tile2},
\hbox{\tt tile1 $\not = \emptyset$} and 
\hbox{\tt tile2 $\not= \emptyset$} as well.
\end{algo} 
\vspace{-12pt}
\grostrait
\vskip 4pt
{\obeylines
\obeyspaces\global\let =\ \tt\parskip=-2pt
   Ltile := theleftmost(tile1,tile2);
   if tile1 = Ltile
     then Rtile := tile2;
     else Rtile := tile1;
   end if;
   Lcursor := chain\_translation(pathroot(Ltile));
   Rcursor := chain\_translation(leftmost(pathroot(Rtile)));
   loop
      measure(distance,Lcursor,RcursoR);
      if Lcursor.next = null then exit; end if;
      if Rcursor.next = null then exit; end if;
      exit when distance > 1;
      Lcursor := Lcursor.next;
      Rcursor := Rcursor.next;
   end loop;
   -- we go out of the loop when distance $\geq$~2
   -- the distance is between Lcursor.next et Rcursor.next
   ladresse := connect (Lcursor,Rcursor);
   return ladresse;
\par}
\demitrait
\vskip 10pt
}

   When the leftmost tile is determined, then the function first computes
the path from {\tt Ltile} to the central cell and then the leftmost path
from {\tt Rtile} to the central cell. In this way, we already know that
the distance between the paths is at most~1 as long as possible. In the
loop which starts form the central tile, the distance between the tiles 
of the path at the current stage is measured by the procedure
{\tt measure} which updates {\tt distance} which outputs the computed distance. 
If one path is completely
traversed or if the distance is now bigger than~1, the loop is completed
and both pointers point at the furthest tile from the central tile
where the distance is at most~1. Then the function {\tt connect} establishes
the necessary path in order to join both remaining parts of the paths in the shortest
way: it is a finite selection of cases in each of which the construction is easy. 
This point raises no difficulty.

The procedure~{\tt measure} is detailed by Algorithm~\ref{measure}. It takes into
account that the leftmost path which joins the central cell to the rightmost tile
may lie on the left-hand side of the path joining the leftmost tile. This is why we
have this selection of cases into three of them denoted by {\tt equal}, when the
distance is~0, {\tt normal} when the distance is~1 and the tile on {\tt Lmark}
is on the left-hand 
side of the tile on {\tt Rmark}, and {\tt opposite} when
the distance is~1 and the tile on 
{\tt Lmark} is on the right-hand side of the
tile on {\tt Rmark}. Such cases do happen. Note that the case {\tt opposite}
is structurally very similar to the case {\tt normal}. The only difference
is that in the former case, we look at {\tt Rmark.status} while in the
latter we look at {\tt Lmark.status}. The reason is that in both cases, we have
to consider the status of the rightmost tile.

\vtop{
\begin{algo}\label{pathroot}
\small
The computation of the path from the root to a node of the tree.
shortest path. Given \hbox{{\tt tile} $=$ $(${\tt tile}$(0)$,{\tt tile}$(1))$},
the number of the sector and the number of the node respectively.
Also, \hbox{{\tt representation} = $[${\tt tile}$(1)]$} is an array of $0$'s and $1$'s
indexed from~$0$. The index {\tt cursor} starts from~$1$ and it is on the
lowest digit.
\end{algo} 
\vspace{-12pt}
\grostrait
\vskip 4pt
{\obeylines
\obeyspaces\global\let =\ \tt\parskip=-2pt
   stage := new T\_pairs;
   stage.next := thepath;
   stage.ingate := 1;
   stage.outgate := 0; -- characterizes an end
   thepath := letape;
   while cursor < representation'last
   loop
      stage := new T\_pairs;
      stage.next := thepath;
      stage.ingate := 1;
      if cursor+1 > representation'last
        then
           stage.outgate := 4 + representation(cursor);
        else
           stage.outgate := 4 - representation(cursor+1)
                           + representation(cursor);
           if representation(cursor+1) = 1
             then if cursor+2 <= representation'last
                    then representation(cursor+2) := 1;
                  end if;
           end if;
      end if;
      thepath := letape;
      cursor := cursor+2;
   end loop;
   --  finalization :
   stage := new T\_pairs;
   -- inversion
   stage.sortie := tile(0);
   stage.entree := 0;
   stage.next := thepath;
   thepath := stage;
   return thepath;
\par}
\demitrait
\vskip 10pt
}

    This allows us to briefly mention that the function {\tt leftmost}
works in a similar way, also based on the computation of the distance
between the given path and the constructed one which should be the leftmost one.
During the construction, the algorithm tries to keep the distance between the
current tile on the given path and the current tile of the constructed path
exactly equal to~1. As long as this is possible, the algorithm goes on in this
way. If it can hold the condition until the last connection to the target of the
path, we are done. But it may happen, and this indeed does happen, that
at the next step after the current path, the distance must be at least~2. This 
means that the constructed path went to much to the left and that it is needed
to resume the computation 
\ligne{\hfill}
\vtop{
\vspace{-20pt}
\begin{algo}\label{measure}
\small
The computation of the distance between two tiles, assuming that
the distance between their fathers is at most~$1$.
Given {\tt Lmark},{\tt Rmark}, two pointers on the considered paths.
\end{algo} 
\vspace{-12pt}
\grostrait
\vskip 4pt
{\obeylines
\obeyspaces\global\let =\ \tt\parskip=-2pt
   distance0 := distance;
   case side is
   when equal => -- distance = 0
      if Lmark.ingate = 0 then -- initialisation
        maxim := maxi(Lmark.outgate,Rmark.outgate);
        minim := mini(Lmark.outgate,Rmark.outgate);
        ecart1 := maxim - minim;
        ecart2 := minim + 7 - maxim;
        ecart := mini(ecart1,ecart2);
      else -- ordinary situation: distance0 = 0
        if (Lmark.outgate /= 0) and (Rmark.outgate /= 0) then 
          distance := maxi(Lmark.outgate,Rmark.outgate)
                      - mini(Lmark.outgate,Rmark.outgate);
        else distance := distance0;
        end if;
      end if;
      if Lmark.outgate /= Rmark.outgate then 
        if Lmark.outgate = theleftmost(Lmark.outgate,
                                       Rmark.outgate) then 
          side := normal; 
        else side := opposite; 
        end if;
      end if;
   when normal => -- distance0 = 1
      if (Lmark.outgate /= 0) and (Rmark.outgate /= 0) then
        distance := 5 - Lmark.outgate + Rmark.outgate - 3;
        if Rmark.status = white then 
          distance := distance+1; 
        end if;
        if distance = 0 then side := equal; end if;
      else distance := distance0;
      end if;
   when inverse =>
      if (Lmark.outgate /= 0) and (Rmark.outgate /= 0) then
        distance := 5 - Lmark.outgate + Rmark.outgate - 3;
        if Lmark.status = white then 
          distance := distance+1; 
        end if;
        if distance = 0 then side := equal; end if;
      else distance := distance0;
      end if;
   end case;
\par}
\demitrait
\vskip 10pt
}

\noindent
from a previously reached tile. If it were needed
to go back to the central time each time the computation has to be resumed,
the algorithm would be quadratic. 

Fortunately, a careful analysis of the construction
shows that it is possible to find key tiles so that each time we have
to resume the computation, we have to go back to the last fixed key tiles so that
adding all the traversed parts of the path lead to a total length which is at most
twice the length of the given path. The length of this part of the program
exceeds the room for this paper. However, the overall computation of the
function {\tt leftmost} is still linear in time with respect to the length of the
given path.

   As a last auxiliary function, we just remind the algorithm for the Poisson
generator, taking from~\cite{knuth}. We can rewrite it as follows:

\vtop{
\begin{algo}\label{poisson}\small
The Poisson generator for an integer valued function.
Here, {\tt random} is a uniform random integer-valued variable in the  
range $0$..{\tt p\_rand}. The function {\tt random} is an algorithmic random
generator. It can be constructed as indicated also in~{\rm\cite{knuth}}.
\end{algo}
\vspace{-12pt}
\grostrait
\vskip 4pt
{\obeylines
\obeyspaces\global\let =\ \tt\parskip=-2pt
      compte : integer := 0;
      pois   : double := 1.0;
   begin
      loop
         pois := pois*double(random)/double(p\_rand);
         exit when pois < exp(-lambda);
         compte := compte+1;
      end loop;
      return compte;
   end;
\par}
\demitrait
\vskip 10pt
}

\subsection{Implementing the protocol}
\label{scenar}

   The simulation is controlled by the procedure {\tt execute}, see
Algorithm~\ref{exec}. As there are many variables to collect partial results
for later analysis, we use records in order to make the program more readable.
The partial results concern the various kinds of messages, so that the
fields of the records are {\tt public}, {\tt reply}, {\tt write}, 
{\tt nonpublic}, {\tt erase}.

   The function {\tt init\_config} initializes the table {\tt space}. In particular,
for each tile, it provides the information about the number of the tile, its sector,
as well as the similar information for its seven neighbours. It also computes
the table {\tt associates} of the numbers of the sides of tile in the neighbour
sharing this side.

   In Algorithm~\ref{exec}, note that we perform addition on records thanks
to the facility given by $ADA$ to overload the usual signs of operation '+',
'$-$', '$\times$' as well as relations, '$=$' and '$>=$' as an example.

   We develop the function {\tt transition} in Algorithm~\ref{transit}.

   The actual content of the function {\tt transition} is performed by the
procedure {\tt action\_in} which implements the exact choices of the simulation,
see Algorithm~\ref{action}.

   As indicated in Section~\ref{scenario}, public messages are sent and conveyed at
odd times. Due to the copying process from{\tt stack0} onto {\tt stack1} on which 
the execution is based, the information about a public message has to be copied on 
{\tt stack1} at even times: otherwise, the message would be erased. This is what
the procedure {\tt replicate} performs. The same procedure is also used for the
erasing messages during the delay they observe at the tile which emitted a public
message. Of course, at each replication, the delay is decreased by~1, until the
delay reaches~1. At this moment, it is sent to catch the emitted message.

\vtop{
\vspace{-4pt}
\begin{algo}\label{exec}\small
The procedure {\tt execute}.
\end{algo}
\vspace{-12pt}
\grostrait
\vskip 4pt
{\obeylines
\obeyspaces\global\let =\ \tt\parskip=-2pt
   auxil := init\_config;
   collect(auxil,0,collect\_at\_t,nb\_max\_msg\_t);
   totals := collect\_at\_t;
   the\_max\_msg := nb\_max\_msg\_t;
   themax\_at\_t := collect\_at\_t;
   for t in 1..duration
   loop
      space := transition(auxil,t);
      auxil := space;
      collect(auxil,t,collect\_at\_t,nb\_max\_msg\_t);
      totals := totals + collect\_at\_t;
      themax\_at\_t := maxi(themax\_at\_t,collect\_at\_t);
      the\_max\_msg := maxi(the\_max\_msg,nb\_max\_msg\_t);
   end loop;
\par}
\demitrait
\vskip 10pt
}

\vtop{
\vspace{-4pt}
\begin{algo}\label{transit}\small
The function {\tt transition}.
\end{algo}
\vspace{-12pt}
\grostrait
\vskip 4pt
{\obeylines
\obeyspaces\global\let =\ \tt\parskip=-2pt
   -- central tile:
   action\_in(0,0);
   -- the other tiles:
   for sect in 1..space'last(1)
   loop
      for i in 1..space'last(2)
      loop
         action\_in(sect,i);
      end loop;
   end loop;
   return copy(new\_space);
\par}
\demitrait
\vskip 10pt
}

   Let us describe this procedure with more details. 
   Let $\mu$ be a public message issued at time~$2t$$-$1, with $t>0$ by the procedure
{\tt send}. At the same time, the procedure {\tt send} creates an erasing message
$\mu_e$ with the same number as~$\mu$. In {\tt send}, the Poisson random generator
is called with the parameter {\tt poisson\_radius} in order to define the radius
of propagation of~$\mu$. This initializes the field {\tt wait} attached to~$\mu_0$.
By construction, the radius is always positive. Note that there is no condition on
the time for the management of an erasing message. As long as its delay is greater
than~1, the erasing message remains in the tile~$T$ where it was created. When the
delay is~1, it is sent to the neighbours with its field {\tt wait} set to~0.
The procedure {\tt convey} transmits the erasing message as its delay is now
always~0, called at each time by {\tt action\_in}, see Algorithm~\ref{action}.

\vtop{
\vspace{-4pt}
\begin{algo}\label{action}\small
The procedure {\tt action\_in}.
\end{algo}
\vspace{-12pt}
\grostrait
\vskip 4pt
{\obeylines\leftskip 0pt
\obeyspaces\global\let =\ \tt\parskip=-2pt
   cursor := space(sect,i)(0).message\_stack;
   while (cursor /= null)
   loop
      case cursor.the\_type is
      when public =>
         if (time mod 2) /= 0 then
           if cursor.direct = null then
             send(from => (sect,i), cursor => cursor);
           else convey (from => (sect,i), cursor => cursor);
           end if;
         else -- even time:
           replicate(cursor, ontile => tile1);
           if poissonrandom(poisson\_reply) > 0 then
             reply(from => (sect,i), place => cursor,
                            num => cursor.number);
           end if;
         end if;
      when nonpublic =>
         convey (from => (sect,i), cursor => cursor);
      when erasing => -- cursor.direct = null, always 
         if cursor.wait = 1 then
           send(from => (sect,i), cursor => cursor);
         elsif cursor.wait = 0 then
           convey (from => (sect,i), cursor => cursor);
         else -- cursor.wait > 1
           replicate(cursor, ontile => tile1);
           tile1.last\_message.wait := cursor.wait-1;
         end if;
      end case;
      cursor := cursor.next;
      compte := compte+1;
   end loop;
   if poissonrandom(poisson\_write) > 0 then
     write (from => (sect,i));
   end if;
   if (time mod 2) = 0 then 
     if poissonrandom(poisson\_public) > 0 then
       init\_cell(sect,i,new\_space,public);
     end if;
   end if;
\par}
\demitrait
\vskip 10pt
}

It is not difficult to see that if~$\mu$ is sent at the time~$t$ and $\mu_e$
is sent at the time~$t$+$r$, then $\mu$ and~$\mu_e$ are at the same tile at the
time~$t$+$2r$.

   More precisely, $\mu$ and~$\mu_e$ are present on the same tile at an even time:
$t$+$2r$ has the same parity as $t$ which is odd and when $\mu$ is sent,
but $\mu_e$ is sent when $\hbox{\tt wait} = 1$. And so, the coincidence is detected 
by the procedure {\tt convey} at the time $t$+$2r$+1: at this time,
$\mu$, which reached the tile~$\tau$ at~$t$+$2r$ is still there at $t$+$2r$+1.
The situation is given in a schematic way by Table~\ref{erasing} and it is 
illustrated by Figure~\ref{fig_erasing}.

\vtop{
\vspace{-4pt}
\begin{algo}\label{erasing}\small
The procedure {\tt action\_in}.
\end{algo}
\vspace{-12pt}
\grostrait
\vskip 4pt
{\obeylines\leftskip 0pt
\obeyspaces\global\let =\ \tt\parskip=-2pt
--
-- sector      1   1   1       5   5   5
-- tile        7   3   1   0   1   9  24    
-- wait        0   1   2   3   4   5   6       
--
--        0    X
--        1        X                              6 
--        2        X                              5
--        3            X                          4
--        4            X                          3
--        5                X                      2
--        6    o           X                      1
--        7        o           X                  0
--        8            o       X                  0
--        9                o       X              0
--       10                    o   X              0
--       11                        o   X          0
--       12                            Z          0
--
\par}
\demitrait
\vskip 10pt
}

\vtop{
\vspace{-25pt}
\centerline{
\mbox{\includegraphics[width=300pt]{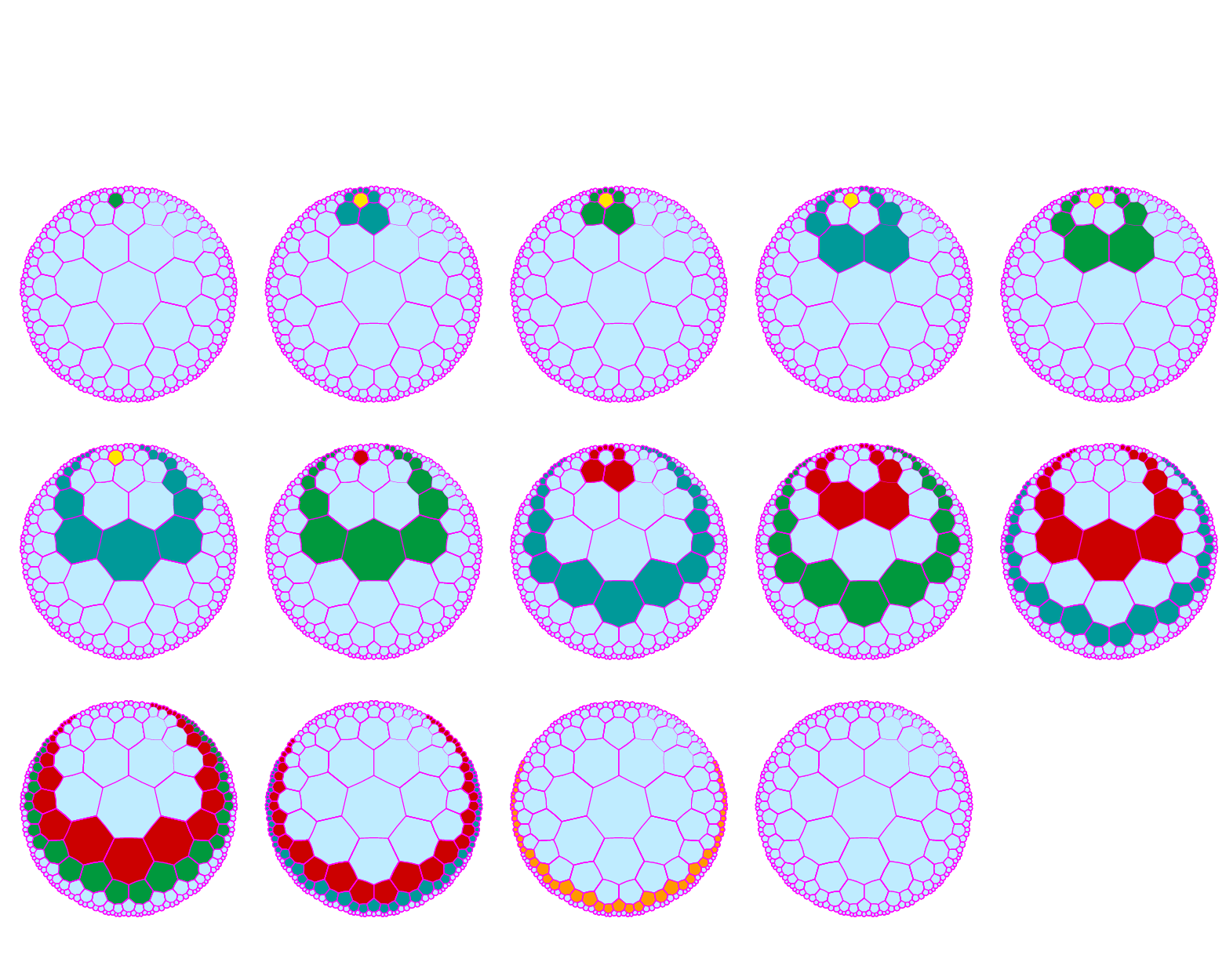}}
}
\vspace{-5pt}
\begin{fig}\label{fig_erasing}
\small
The emission of a public message and of its erasing signal.
\end{fig}
}
\vskip 7pt
   In the program, the cancellation of a message reached by its erasing signal
is performed by the procedure {\tt convey}. When dealing with a public message,
the first task of the procedure is to scan the stack in order to possibly detect
an erasing signal with the same number as the message. When this happens, the
message, which is scanned in {\tt stack0} is simply not copied onto {\tt stack1}. 
When the procedure examine an erasing signal, it performs a similar scanning:
if a public message bears the same number as the signal, the signal is not copied
onto {\tt stack1}. As the cancellation has to occur when both signals are present,
this simple way is enough to perform this action and there is no need
to connect the two decisions of not copying the information onto 
{\tt stack1}. This is guaranteed by the disjunction between {\tt stack0} and 
{\tt stack1}.

\section{\Large The experiment}
\label{experiment}

   The experiment was performed by the running the program on a simple laptop.
The laptop is a {\tt Lenovo} one, with two Intel processors, both working
at 2~GHz, and Linux Mandriva as operating system. The used $ADA$-compiler 
belongs to the {\tt gnu}-family, version 4.4.1. In the first sub-section, we
describe the experiment and we give an account of the results. In the second
sub-section, we give an interpretation of the results.

\subsection{Description of the experiment and of the results}
\label{rawdata}

   The program was run for six value of the depth of the Fibonacci tree,
ranging from~5 to~10. 
Denote by $\cal S$ the observation space.
The size of~$\cal S$ is defined by the depth of the Fibonacci tree which spans the
seven sectors displayed around the central cell. We consider the tiles whose
level in the tree is at most {\tt depth} which takes values in [5..10] in our
experiments. This means that $\cal S$ contains 1625 tiles when 
$\hbox{\tt depth} = 5$ and 200593 tiles when $\hbox{\tt depth} = 10$. 
Increasing the depth by~1 means multiplying the number of tiles by 
a coefficient which quickly tends to
$\displaystyle{{3+\sqrt5}\over2}\approx 2.618034$. This factor of
a bit more than 2.6, can be observed in Table~\ref{statnb1} which indicates
the number of tiles of~$\cal S$ for the different values of {\tt depth}. 

The second parameter which we also changed
is the radius of propagation of the public messages. As indicated in 
Section~\ref{scenario}, the radius is an integer valued random variable following
a Poisson law. The coefficient is fixed to~5{} in one series of experiments and 
to~10{} in the second one. The range of the variable is in mean this
coefficient. However, the value may range from~0 to twice the coefficient with,
from time to time, bigger values. Table~\ref{statnb1} indicates the number of messages 
emitted in the observed area under these conditions. The remaining parameters
are the following. Each tile is given the possibility to 
emit a message. Again we assume that the probability of such an event is 
given by a Poisson law. The coefficient is 0,005 for a public message, taking into
account that such messages are emitted at odd times only. Also, the tiles which 
are on the border of the space and which are the more numerous, more than 60\%{}  
of the overall number of tiles in~$\cal S$, are given an additional possibility 
with again a Poisson law whose coefficient is 0.0025. This is also to reflect the 
possibility for the tiles of~$\cal S$ to receive messages from outside~$\cal S$.
Now, for private messages, they are caused either as a reply to a public message,
or as a single message sent to a particular tile via the consultation of the
directory. These events are also following a Poisson law and the coefficient
is 0.0025 for the reply to a public message, and 0.001 for the consultation
of a directory. In the case of a message sent after consulting the directory,
it is assumed that the coordinate of the tile given by the directory falls 
within~$\cal S$ and that both numbers constituting the coordinate are uniformly 
distributed in their respective ranges.

\newdimen\thelarge\thelarge=45pt 
\def\lignure #1 #2 #3 #4 #5 #6 #7 {%
\ligne{\hfill
\hbox to \thelarge{\hfill#1\hfill}
\hbox to \thelarge{\hfill#2\hfill}
\hbox to \thelarge{\hfill#3\hfill}
\hbox to \thelarge{\hfill#4\hfill}
\hbox to \thelarge{\hfill#5\hfill}
\hbox to \thelarge{\hfill#6\hfill}
\hbox to \thelarge{\hfill#7\hfill}
\hfill}
}

\newdimen\thelargebis\thelargebis=45pt 
\newdimen\thestart\thestart=60pt 
\def\lignurebis #1 #2 #3 #4 #5 #6 {%
\ligne{\hskip 10pt
\hbox to \thestart{\hskip 5pt#1\hfill}
\hbox to \thelargebis{\hfill#2\hfill}
\hbox to \thelargebis{\hfill#3\hfill}
\hbox to \thelargebis{\hfill#4\hfill}
\hbox to \thelargebis{\hfill#5\hfill}
\hbox to \thelargebis{\hfill#6\hfill}
\hfill}
}

\vtop{
\begin{tab}\label{statnb1}
\small The number of tiles and the total number of messages. The number after 
{\tt sent} is the radius of propagation. The time lines indicate the number of
iterations during which the program was executed. Under the line indicating the 
time for radius~$5$, we indicate the ratio between the number of messages
for consecutive depths as long as the overall duration is the same.
The lines {\tt mean} indicate the mean of the number of messages up to~$t$
divided by~$t$ for $t\in[1..T]$ where $T$ is indicated by the line {\tt time}.
\end{tab}
\vspace{-12pt}
\grostrait
\vspace{4pt}
\lignure {depth} 5 6 7 8 9 {10}
\lignure {tiles} {1625} {4264} {11173} {29261} {76616} {200593}
\lignure {sent, 5} {1101} {3308} {8636} {21797} {49295$^*$} {60453$^*$}
\lignure {time, 5} {168} {168} {168} {168} {142} {69}
\lignurebis {ratio} {3.00454} {2.61064} {2.52396} {} {}
\lignure {mean, 5} {6.81949} {19.13115} {50.10053} {128.31127} {342.79960} {877.83673}
\lignurebis {ratio} {2.80536} {2.61879} {2.56108} {2.67162} {2.56079}
\lignure {sent, 10} {2173} {8289} {13687$^*$} {13784$^*$} {19167$^*$} {30164$^*$}
\lignure {time, 10} {168} {168} {92} {41} {30} {24}
\lignure {mean, 10} {11.21332} {40.57043} {101.35430} {197.08219} {405.77815} 
{965.53752}
\lignurebis {ratio} {3.618057} {2.50356} {1.94449} {2.058929} {2.37947}
\vspace{-2pt}
\demitrait
\vskip 10pt
}

\vtop{
\begin{tab}\label{statnb2}
\small The number of tiles and the total number of messages at time~$24$. 
The conventions are those of Table~{\rm\ref{statnb1}}. Under each line
indicating the number of messages sent at time~$24$, we have the ratio
between two consecutive numbers.
\end{tab}
\vspace{-12pt}
\grostrait
\vspace{4pt}
\lignure {depth} 5 6 7 8 9 {10}
\lignure {tiles} {1625} {4264} {11173} {29261} {76616} {200593}
\lignure {sent, 5} {169} {423} {1158} {3017} {7888} {20556}
\lignurebis {ratio} {2.50296} {2.73759} {2.60535} {2.61452} {2.60599}
\lignure {sent, 10} {204} {582} {1538} {4509} {11413} {30164}
\lignurebis {ratio} {2.85294} {2.64261} {2.94129} {2.53116} {2.64295}
\vspace{-2pt}
\demitrait
\vskip 10pt
}

Another important feature regarding private messages is that in the experiment,
it was assumed that once a communication has started it goes on endlessly: if 
$A$~replies to a message sent by~$B$, either public or private, then $B$~replies
to~$A$ which again replies to~$B$ and this process goes on periodically. Moreover,
the reply was always assumed to be immediate. 

Table~\ref{statnb1} indicates the number of messages issued during the whole
time of the simulation, measured by the number of iterations of the 
procedure {\tt execute}. The number of iterations is also given by the table.
It can be noticed that this number is always 168 for small values. This number 
was fixed for the experiment and it can be noticed that \hbox{168 = 7$\times$24}.
For a fixed value of the mean radius of propagation, we notice that the
number of iteration becomes lower and lower. This is a limitation caused by
the system and the machine under which the program was run. It can be noticed that 
the decay of the number of the iterations corresponds to the increase of the
number of tiles. Accordingly, this alters the number of messages which were issued.

Table~\ref{statnb1} indicates the ratio between the numbers of messages sent
when the radius of propagation is~5 and when the depth of the Fibonacci tree is~5,
6 and~7 as for these depths, we have the same duration~168. In Table~\ref{statnb2},
we indicate the overall number of sent messages at time~24 as we have data for each
depth of the experiment. This allows us to compute the ratio between numbers 
associated to consecutive depths. 

\vtop{
\begin{tab}\label{statmax1}
\small The number of tiles and the maximal number of messages passing through 
a tile at a time during the interval of observation.
The number after {\tt max} is the radius of propagation. The rest of the conventions
are those of Table~{\rm\ref{statnb1}}.
\end{tab}
\vspace{-12pt}
\grostrait
\vspace{4pt}
\lignure {depth} 5 6 7 8 9 {10}
\lignure {tiles} {1625} {4264} {11173} {29261} {76616} {200593}
\lignure {max, 5} {18} {39} {58} {104} {232$^*$} {192$^*$}
\lignurebis {ratio} {2.16667} {1.48718} {1.79310} {} {}
\lignure {time, 5} {168} {168} {168} {168} {142} {69}
\lignure {max, 10} {63} {169} {204$^*$} {197$^*$} {315$^*$} {694$^*$}
\lignure {time, 10} {168} {168} {92} {41} {30} {24}
\vspace{-2pt}
\demitrait
\vskip 10pt
}

\vtop{
\begin{tab}\label{statmax2}
\small The number of tiles and the maximal number of messages at time~$24$. 
The conventions are those of Table~{\rm\ref{statnb1}}. Under each line
indicating the maximal number of messages passing through a tile at time~$24$, 
we have the ratio between two consecutive numbers.
\end{tab}
\vspace{-12pt}
\grostrait
\vspace{4pt}
\lignure {depth} 5 6 7 8 9 {10}
\lignure {tiles} {1625} {4264} {11173} {29261} {76616} {200593}
\lignure {sent, 5} {11} {16} {25} {34} {54} {91}
\lignurebis {ratio} {1.45455} {1.56250} {1.36000} {1.58824} {1.68519}
\lignure {sent, 10} {17} {30} {61} {140} {315} {694}
\lignurebis {ratio} {1.76471} {2.03333} {2.295082} {2.25000} {2.20317}
\vspace{-2pt}
\demitrait
\vskip 10pt
}

Table~\ref{statnb1} also reports another measurement performed by the program:
if $n_t$ is the number of messages emitted up to the time~$t$ with $t\in [1..T]$,
where $T$ is the duration of the experiment, {\it i.e.} the number of iterations,
then the lines {\tt mean} give the mean value of the numbers 
$\displaystyle{{n_t}\over t}$. These values are computed when the propagation
radius is~5 and when it is~10.
 
Another interesting information is the maximal number of messages passing through
a tile. The data are given in Table~\ref{statmax1}, in the same conditions as
in Table~\ref{statnb1}.

The data are summarized in Tables~\ref{statmax1} and~\ref{statmax2}.

Below, Tables~\ref{statcompo1} and~\ref{statcompo2} give an information on
the decomposition of the number of emitted messages between the public messages
and the private ones

\newdimen\thelargea\thelargea=45pt 
\def\lignurea #1 #2 #3 #4 #5 #6 {%
\ligne{\hfill
\hbox to \thelargea{\hfill#1\hfill}
\hbox to \thelargea{\hfill#2\hfill}
\hbox to \thelargea{\hfill#3\hfill}
\hbox to \thelargea{\hfill#4\hfill}
\hbox to \thelargea{\hfill#5\hfill}
\hbox to \thelargea{\hfill#6\hfill}
\hfill}
}

\vtop{
\begin{tab}\label{statcompo1}
\small This table is a refinement of Table~{\rm\ref{statnb1}}. It indicates
how the number of messages emitted in~$\cal S$ are distributed between the
public and the private messages and, among the latter ones, between the replies
to a public message or a direct message to a single tile via the directory.
This table gives the data for radiuses~$5$ and~$10$ for the propagation of 
the public messages, upper and lower halves of the table, respectively.
\end{tab}
\vspace{-12pt}
\grostrait
\vspace{4pt}
\lignurea {depth} {time} {public} {reply} {write} {total}  
\vspace{7pt}
\lignurea {5} {168} {636} {211} {254} {1101}  
\lignurea {ratio} {}    {0.577} {0.192} {0.231} {}  
\lignurea {6} {168} {1840} {783} {685} {3308}
\lignurea {ratio} {}    {0.556} {0.237} {0.207} {}  
\lignurea {7} {168} {4669} {2285} {1682} {8636}
\lignurea {ratio} {}    {0.541} {0.264} {0.195} {}  
\lignurea {8} {168} {11982} {5295} {4520} {21797}
\lignurea {ratio} {}    {0.550} {0.243} {0.207} {}  
\lignurea {9} {142} {27099} {12488} {9708} {49295}
\lignurea {ratio} {}    {0.550} {0.253} {0.197} {}  
\lignurea {10} {69} {34536} {13467} {12450} {60453}
\lignurea {ratio} {}    {0.571} {0.223} {0.206} {}  
\vspace{-2pt}
\demitrait
\vspace{4pt}
\lignurea {5} {168} {654} {1281} {238} {2173}    
\lignurea {ratio} {} {0.301} {0.589} {0.110} {}  
\lignurea {6} {168} {1791} {5848} {650} {8289}   
\lignurea {ratio} {} {0.217} {0.705} {0.078} {}  
\lignurea {7} {92}  {2472} {10279} {936} {13687}   
\lignurea {ratio} {} {0.181} {0.751} {0.068} {}  
\lignurea {8} {41}  {3094} {9603} {1087} {13784}
\lignurea {ratio} {} {0.224} {0.697} {0.079} {}  
\lignurea {9} {30}  {6107} {10937} {2123} {19167}
\lignurea {ratio} {} {0.318} {0.571} {0.111} {}  
\lignurea {10} {24} {12935} {12954} {4275} {30164}
\lignurea {ratio} {} {0.429} {0.429} {0.142} {}  
\vspace{-2pt}
\demitrait
\vskip 10pt
}

\newdimen\thelargea\thelargea=45pt 
\def\lignureb #1 #2 #3 #4 #5 {%
\ligne{\hfill
\hbox to \thelargea{\hfill#1\hfill}
\hbox to \thelargea{\hfill#2\hfill}
\hbox to \thelargea{\hfill#3\hfill}
\hbox to \thelargea{\hfill#4\hfill}
\hbox to \thelargea{\hfill#5\hfill}
\hfill}
}

\noindent
and, among the latter, between replies to a public message
and direct messages to a tile whose coordinates are delivered by the directory.

As for Table~\ref{statnb1}, Table~\ref{statcompo1} gives the information for
each area of~$\cal S$ defined by the depth of the Fibonacci tree. The upper half
of the table concerns a propagation of the public messages characterized by
radius~5, while, the lower half concerns radius~10.

\vtop{
\begin{tab}\label{statcompo2}
\small This table is a refinement of Table~{\rm\ref{statnb2}}. It indicates
how the number of messages emitted in~$\cal S$ are distributed between the
public and the private messages and, among the latter ones, between the replies
to a public message or a direct message to a single tile via the directory.
All data are taken at iteration~$24$. In the upper half of the table, the radius
of propagation for the public messages is~$5$. In the lower half, the radius 
is~$10$.
\end{tab}
\vspace{-12pt}
\grostrait
\vspace{4pt}
\lignureb {depth} {public} {reply} {write} {total}  
\vspace{7pt}
\lignureb {5} {109} {24} {36} {169}     
\lignureb {ratio} {0.577} {0.192} {0.231} {}  
\lignureb {6} {277} {53} {93} {423}
\lignureb {ratio} {0.556} {0.237} {0.207} {}  
\lignureb {7} {738} {200} {220} {1158}     
\lignureb {ratio} {0.541} {0.264} {0.195} {}  
\lignureb {8} {1856} {504} {657} {3017}
\lignureb {ratio} {0.550} {0.243} {0.207} {}  
\lignureb {9} {5021} {1268} {1599} {7888} 
\lignureb {ratio} {0.550} {0.253} {0.197} {}  
\lignureb {10} {13026} {3181} {4349} {20556}    
\lignureb {ratio} {0.571} {0.223} {0.206} {}  
\vspace{-2pt}
\demitrait
\lignureb {5} {98} {78} {28} {204}   
\lignureb {ratio} {0.480} {0.382} {0.138} {}  
\lignureb {6} {273} {205} {104} {582}
\lignureb {ratio} {0.469} {0.352} {0.179} {}  
\lignureb {7} {672} {614} {252} {1538}
\lignureb {ratio} {0.437} {0.399} {0.164} {}  
\lignureb {8} {1919} {1945} {645} {4509}
\lignureb {ratio} {0.426} {0.431} {0.143} {}  
\lignureb {9} {5010} {4720} {1683} {11413}
\lignureb {ratio} {0.439} {0.414} {0.147} {}  
\lignureb {10} {12935} {12954} {4275} {30164}
\lignureb {ratio} {0.429} {0.429} {0.142} {}  
\vspace{4pt}
\demitrait
\vskip 10pt
}

In Table~\ref{statcompo2}, the data are attached to the same time, defined by
24 iterations. This gives a direct comparison between all the data but it does not
concern a long enough time period.

   We turn to the next sub-section where we try to extract a general information
from these data and from a few other ones we have not the room to give in this
paper.

\subsection{Interpretation}
\label{interpret}

   Several conclusions can be drawn from the results presented in 
Subsection~\ref{rawdata}.

   The first one concerns the ratios between the number of messages
for consecutive depths of the Fibonacci trees. These ratios are close to the ratio
between the area of~$\cal S$ for consecutive values of the depth of the spanning
tree. It seems that we may conclude that these experimental data support 
Assumption~\ref{proparea}. 

\begin{hypo}\label{proparea}
For any $t$, the number of messages issued at the time~$t$ in~$\cal S$ is
proportional to the number of tiles belonging to~$\cal S$.
\end{hypo}

   Indeed, the coefficient of the Poisson law is a kind of mean of the random
variable indicating whether a message is sent or not. According to the homogeneous
nature of the space and as the decision of one tile is independent from that of
its neighbours, it can be expected that the observed number of issued messages
is proportional to the area. This conclusion is strengthened by the following
consideration. As the actual radius of propagation of the public messages is
bounded, the contribution of far tiles is ruled out, starting from a certain
distance from a tile and this distance can be uniformly bounded for all the tiles.
The values of the radius exceeding say twice the mean of the radius as a random
variable which we assumed to follow a Poisson law can be considered as an event
of very small probability so that an infinite repetition of exceptionally long 
radiuses can be considered as an event of probability~0. The same remark apply if we
relax a bit the conditions on the coordinates provided by the directory. We may
assume that the number of the sector is uniformly distributed and that the number
of the tile in the tree is an integral random variable following a Poisson law
with a radius of the same size as that of~$\cal S$. And so, relaxing a bit the
condition on the directory as just suggested does not alter the argument in
favor of Assumption~\ref{proparea}.
\vskip 10pt

No clear statement can be inferred from 
Tables~\ref{statmax1} and~\ref{statmax2},
except the fact that the maximal number of messages passing through~$\cal S$ seems
to be increasing with the time. This can also be seen on the experiment for each
depth: as the number of iteration increases, the maximal number of messages
passing at a tile also increases.

   It is also interesting to look at where the maximal number of messages appear.
We have not the room to give the relevant information computed by the program.
For each iteration, the program indicates a tile at which the number of
passing messages is maximal. It also splits the information in looking at where
is obtained the maximal number of passing public messages or private messages
replying to a public message or private messages written to a single tile.
It is interesting to notice that most often the tiles are not the same for the
different kinds of messages. Also, the position of this maximum is generally
the central tile or one of its neighbours. However, for the emission of the
pubic message, the maximal number of passages at a tile may be obtained a bit 
further from the central tile, while the replies to these messages seem
to be maximal most often at the central tile or its immediate neighbours.

   Tables~\ref{statcompo1} and~\ref{statcompo2} show a very interesting difference
between the cases when the radius of propagation of the public messages is~5 and
when it is~10. In both tables, we can see that the proportion of public messages
is higher when the radius is~5. This is particularly striking in 
Table~\ref{statcompo1}, but it is already noticeable in Table~\ref{statcompo2}
showing that this difference appears quickly and that it tends to increase a
bit with the time. It is also interesting to see that the relative 'loss'
of the public messages 'benefit' to their replies. Indeed, in both tables, 
the proportion of direct messages to a single tile is not improved when the
depth of the Fibonacci tree is increasing. Of course, the Poisson coefficient
for triggering a public message is 0.005, while that of a reply is 0.0025
and that of a direct private message is 0.001. However, the public messages are 
triggered at odd times only and the replies occur only at even time while
the direct private messages can be sent at any time. There should be no big
difference between direct private messages and replies to a public message.
In fact the explanation lies in the geometry of the space. Indeed, the reply
is proposed to any tile visited by the propagation wave which covers all the tiles
within the radius fixed at the time when the public message was emitted. 
This additional solicitation explains the importance of the replies.

   At this point, we can also indicate why we have chosen a Poisson law to model
this message system. The reason is that we have to take into account the geometry
of the space. A uniform distribution would give much more weight to distant tiles
by the simple fact that their number increases exponentially: the farther they
are the more message they would send to the centre. This is also the reason why
we decided to limit the propagation of the public messages. If no limitation
would be put, the number of messages received at any point would grow exponentially
with time by the just mentioned argument. Accordingly, the limitation restricts
this possibility. Now, the Poisson law also gives the possibility to obtain big
values with respect to the mean value. Simply these extremal are very rare, the
more rare they are higher values.

\section{\Large Conclusion}
\label{conclusion}

   It is the place here to discuss how these results can be
interpreted in a more qualitative way. The number of iterations suggests that
the unit of time is an hour. The tiles can be interpreted either as individuals
or as groups of individuals in a given constant area, the one defined by the
area of a tile. Remember that in the space we consider, all tiles have the same
area. The limitation of the public messages can be interpreted as a natural
limit due to the conditions in which the message is sent and also depending
on the intentions of the sender.

   Two important points should be noted. The first one is the property of the
public messages to cover all the tiles of a given area for each tile to receive 
the message once exactly. The second point is the mechanism to limit the
propagation of a public message. This mechanism needs no centralization. It is
monitored by the sender and, a priori, each tile can be a sender. To be a sender
is defined by a probability which is the same for every tile.
The third point is that in some sense, the indicated scenario is a worse
case one with respect to the traffic load supported by each tile. Indeed,
the fact that once a communication is established between two tiles goes on
endlessly contributes to increase the traffic with the time. There is room here
to tune the modeling by introducing various ways to delay answers or to 
limit the number of contacts of a tile with others: here also we could consider
that this number is a Poisson random variable whose mean can be fixed uniformly
or depending on other criteria which we have not considered here.
A last point is the possible improvement of the program in order to obtain more
data and to go further in the exploration of the simulation space. Note that
the file which records all the communications when the depth is~5 and the
radius of propagation is~10 and the number of iterations is~168 has a size
of around 164 megabytes. As mentioned in Section~\ref{experiment}, increasing the 
depth by~1 multiplies the area 
by around 2.618. Accordingly, the depth which
defines the area of~$\cal S$ cannot be extent very much. Already depth~10
with radius~10 requires a machine more powerful than a simple laptop.

    We are convinced that there is further work ahead to better analyze the 
data already obtained, to improve the program in order to go further in the 
exploration of the simulation space. There is also room to tune the basic
parameters in order to get a picture closer to real networks as, for instance,
social networks.





\end{document}